\documentstyle[a4,aps,floats,epsf,psfig,fleqn]{revtex}
\DeclareFontFamily{U}{msb}{}
\DeclareFontShape{U}{msb}{m}{n}{
<5><6><7><8><9> gen *msbm <10><10.95><12><14.4><17.28><20.74><24.88>msbm10}{}
\DeclareSymbolFont{AMSb}{U}{msb}{m}{n}
\DeclareMathSymbol{\AAA}{\mathbin}{AMSb}{'101}
\DeclareMathSymbol{\BBB}{\mathbin}{AMSb}{'102}
\DeclareMathSymbol{\CCC}{\mathbin}{AMSb}{'103}
\DeclareMathSymbol{\DDDD}{\mathbin}{AMSb}{'104}
\DeclareMathSymbol{\EEE}{\mathbin}{AMSb}{'105}
\DeclareMathSymbol{\FFF}{\mathbin}{AMSb}{'106}
\DeclareMathSymbol{\GGG}{\mathbin}{AMSb}{'107}
\DeclareMathSymbol{\HHH}{\mathbin}{AMSb}{'110}
\DeclareMathSymbol{\III}{\mathbin}{AMSb}{'111}
\DeclareMathSymbol{\JJJ}{\mathbin}{AMSb}{'112}
\DeclareMathSymbol{\KKK}{\mathbin}{AMSb}{'113}
\DeclareMathSymbol{\LLL}{\mathbin}{AMSb}{'114}
\DeclareMathSymbol{\MMM}{\mathbin}{AMSb}{'115}
\DeclareMathSymbol{\NNN}{\mathbin}{AMSb}{'116}
\DeclareMathSymbol{\OOO}{\mathbin}{AMSb}{'117}
\DeclareMathSymbol{\PPPP}{\mathbin}{AMSb}{'120}
\DeclareMathSymbol{\QQQ}{\mathbin}{AMSb}{'121}
\DeclareMathSymbol{\RRR}{\mathbin}{AMSb}{'122}
\DeclareMathSymbol{\SSS}{\mathbin}{AMSb}{'123}
\DeclareMathSymbol{\TTTTT}{\mathbin}{AMSb}{'124}
\DeclareMathSymbol{\UUU}{\mathbin}{AMSb}{'125}
\DeclareMathSymbol{\VVVV}{\mathbin}{AMSb}{'126}
\DeclareMathSymbol{\WWW}{\mathbin}{AMSb}{'127}
\DeclareMathSymbol{\XXXX}{\mathbin}{AMSb}{'130}
\DeclareMathSymbol{\YYY}{\mathbin}{AMSb}{'121}
\DeclareMathSymbol{\ZZZZ}{\mathbin}{AMSb}{'132}

\newcommand{\ds}{\displaystyle}

\newcommand{\wa}{{\scriptscriptstyle{\cal W}}}
\newcommand{\oi}{{\scriptscriptstyle{\cal O}}}
\newcommand{\ro}{{\scriptscriptstyle{\cal R}}}
\newcommand{\Wc}{{\scriptstyle 1}}
\newcommand{\Wd}{{\scriptstyle 2}}
\newcommand{\Oc}{{\scriptstyle 3}}
\newcommand{\Od}{{\scriptstyle 4}}
\newcommand{\pha}{{\scriptscriptstyle{\alpha}}}
\newcommand{\phb}{{\scriptscriptstyle{\beta}}}

\newcommand{\kr}{k^r}
\newcommand{\kwr}{k_\wa^r}
\newcommand{\kor}{k_\oi^r}

\newcommand{\Sor}{S_{\oi r}}
\newcommand{\Swi}{S_{\wa i}}
\newcommand{\wid}{R}
\newcommand{\one}{{\bf 1}}
\newcommand{\VV}{{\bf v}}
\newcommand{\GG}{{\bf g}}

\newcommand{\MK}{{\bf K}}

\newcommand{\Ca}{{\rm Ca}}

\newcommand{\ph}{{\phi}}
\newcommand{\EG}{{\Psi}}
\newcommand{\Flu}{{\mbox{\boldmath $\tau$}}}
\newcommand{\Ext}{{g}}
\newcommand{\Exc}{{e}}
\newcommand{\EEx}{{E}}
\newcommand{\BF}{{\bf b}}
\newcommand{\OS}{{\sigma}}
\newcommand{\Mob}{{\Lambda}}
\newcommand{\SIS}{{A}}
\newcommand{\SS}{{S}}
\newcommand{\ST}{{\Sigma}}
\newcommand{\SEF}{{\bf q}}
\newcommand{\SEE}{{h}}
\newcommand{\MT}{{M}}
\newcommand{\MMT}{{\bf m}}
\newcommand{\WI}{{W}}
\newcommand{\TT}{{t}}
\newcommand{\ZZ}{{Z}}
\newcommand{\PP}{{P}}

\newcommand{\KK}{{k}}

\newcommand{\XX}{{\bf x}}
\newcommand{\NAB}{{\mbox{\boldmath $\nabla$}}}

\begin{document}
\draft
\setcounter{page}{0}
\title{Macroscopic Equations of Motion for Two Phase Flow
 in Porous Media}
\author{R. Hilfe$\mbox{\rm r}^{1,2}$}
\address{
$\mbox{ }^1$ICA-1, Universit{\"a}t Stuttgart,
Pfaffenwaldring 27, 70569 Stuttgart\\
$\mbox{ }^2$Institut f{\"u}r Physik,
Universit{\"a}t Mainz,
55099 Mainz, Germany}
\maketitle
\thispagestyle{empty}
\begin{abstract}
The established macroscopic equations of motion for two phase
immiscible displacement in porous media are known to be
physically incomplete because they do not contain the
surface tension and surface areas governing capillary phenomena.
Therefore a more general system of macroscopic equations 
is derived here which incorporates the spatiotemporal variation of
interfacial energies.
These equations are based on the theory of mixtures
in macroscopic continuum mechanics.
They include wetting phenomena
through surface tensions instead of the traditional use of
capillary pressure functions.
Relative permeabilities can be identified in this
approach which exhibit a complex dependence on the
state variables.
A capillary pressure function can be identified in
equilibrium which shows the
qualitative saturation dependence known from experiment.
In addition the new equations allow to describe the spatiotemporal
changes of residual saturations during immiscible displacement.
\end{abstract}
\vspace{.7cm}
PACS: 47.55Mh, 81.05Rm, 61.43Gt, 83.10Lk, (47.55Kf)
\begin{flushright}{\em Physical Review E (1998), in print}\end{flushright}
\newpage
\section{Introduction}
An aquifer or a petroleum reservoir may be seen as
mixture of porous rocks and soil with various gases
and liquids, usually residing at depths that prohibit
precise measurements on the constituents
\cite{dul92,adl92}.
Despite many years of research and despite the fact that the
microscopic laws of fluid dynamics are well known
the prediction of macroscopic multiphase flow in porous media
has not met with success (see \cite{sah93,sah95,hil95d}
for recent overviews).

Microscopic and macroscopic descriptions of multiphase
fluid flow in porous media differ considerably from
each other and both have their own characteristic
problems \cite{dul92,adl92,sah95}.
A microscopic description fails because it is generally
impossible and not interesting to know the detailed
microstructure and flow patterns on the pore scale.
In contrast herewith the generally accepted macroscopic
models are incomplete.
Owing to its widespread importance in applied problems
the transition from the microscale to the macroscale for
two phase flow in porous media has attracted considerable
interest from physicists in recent years
\cite{BK90b,MCR91,hil91d,oxa91,OBFJMA91,hil92a,MFFJ92,FFJM92,BS95a,BKS92,KHT92,KRT92,hil94c,sch95,HS95,FMF96,AAMHSS97}.
Relatively little or no attention has been focussed, however,
on the continuum limit of the proposed microscopic models.
Effective macroscopic equations of motion used by engineers
are based on generalizations of Darcy's law, although every 
practitioner knows that they are incomplete \cite{mar81,dul92}.
Microscopic models by physicists often attempt to predict the 
relative permeabilities and the capillary pressure functions 
for simple microstructures although it is clear that
they are highly nonunique.

Detailed and rigorous derivations of macroscopic equations
from microscopic ones
is usually the last step in understanding a
physical phenomenon.
Experimental observations and tests of phenomenological
macroscopic models usually precede the microscopic 
understanding.
It is therefore of interest to improve the current
macroscopic description which is known to fail even
for simple laboratory systems. \cite{mor91}

Given the need for more predictive
macroscopic theories this paper proposes a new set of
macroscopic equations.
Loosely speaking the traditional theory includes only
mass and momentum balance but fails to incorporate the
energy balance.
Obviously the interfacial energy depends on the interfacial
area and hence the set of macroscopic observables describing
the state of the mixture must be enlarged to include 
surface areas.
Relative permeabilities and capillary pressures must then
be identified within the new framework.
I shall begin the discussion by presenting briefly the
traditional formulation and discussing its problems.
Afterwards the macroscopic balance laws for mass, momentum
and energy will be combined with constraints and constitutive
relations to give new macroscopic equations of motion.
My last section in this paper will discuss the consequences
and the identification of relative permeabilities and
capillary pressures.

\section{Problems with existing macroscopic equations}
Consider simultaneous flow of two incompressible and immiscible
fluid phases denoted generically as water (subscript ${\cal W}$)
and oil (subscript ${\cal O}$) inside a porous medium.
The accepted and widely used macroscopic equations of motion
\cite{dul92,sah95,hil94c,sch74}
are based on mass conservation
\begin{equation}
\begin{array}{rcl}
\ph \frac{\ds \partial \SS_\wa}{\ds \partial \TT} & = & -\NAB \cdot \VV_\wa \\[12pt]
\ph \frac{\ds \partial \SS_\oi}{\ds \partial \TT} & = & -\NAB \cdot \VV_\oi
\end{array}
\label{Em1}
\end{equation}
\begin{equation}
\SS_\wa + \SS_\oi = 1
\label{Em3}
\end{equation}
and the well known generalization of Darcy's law \cite{dul92,sch74}
\begin{equation}
\begin{array}{rcl}
\VV_\wa & = & - \left[
\frac{\ds \MK\kwr }{\ds \mu_\wa}\:
(\NAB \PP_\wa - \rho_\wa g\:\NAB\ZZ) \right] \\[12pt]
\VV_\oi & = & -  \left[
\frac{\ds \MK\kor }{\ds \mu_\oi}\:
(\NAB \PP_\oi - \rho_\oi g\:\NAB\ZZ ) \right]
\end{array}
\label{Em2}
\end{equation}
all of which are assumed to be valid on length scales large compared to a
typical pore diameter.
The variables in these equations are the macroscopic pressure fields
of water and oil, denoted as $\PP_\wa$ and $\PP_\oi$, the water and
oil saturations $\SS_\wa,\SS_\oi$ and the macroscopic (Darcy) velocity
fields $\VV_\wa,\VV_\oi$.
The saturation is defined as the ratio of water (oil) volume to pore 
space volume.
Pressures and saturations are averages over a macroscopic region 
much larger than the pore size, but much smaller than the system size.
Their arguments are the macroscopic space and time variables
$(\XX,\TT)$. 
The porosity $\ph$ is the ratio of pore volume to total volume of the
medium.
The fluid viscosities and densities are denoted by $\mu$ and $\rho$.
The acceleration of gravity is $g$ and the function $\ZZ(\XX)$ is the 
depth function.
In the equations above $\MK$ stands for the
absolute (single phase flow) permeability tensor, $\kwr$ 
is the relative permeability for water, $\kor$ the 
oil relative permeability. 
The equations above are closed by defining the capillary pressure as
\begin{equation}
\PP_c(\XX,\TT) = \PP_\oi(\XX,\TT) - \PP_\wa(\XX,\TT) 
\label{Em4}
\end{equation}
and by postulating for it the constitutive relationship
\begin{equation}
\PP_c(\XX,\TT) = \PP_c(\SS_n(\XX,\TT))
\label{Em5}
\end{equation}
where
\begin{equation}
S_n(\XX,\TT) = \frac{\SS_\wa(\XX,\TT)-\Swi}{1-\Swi-\Sor} 
\label{Em6}
\end{equation}
is a normalized saturation.
The water saturation obeys $\Swi < \SS_\wa < 1-\Sor$ where the two
numbers $0\leq\Swi,\Sor\leq 1$ are two parameters representing the
irreducible water saturation, $\Swi$, and the residual oil
saturation, $\Sor$.
The residual oil saturation gives the amount of oil remaining
in a porous medium after a water flood.
The normalized saturation $\SS_n$ varies between $0$ and $1$
as $\SS_\wa$ varies between $\Swi$ and $1-\Sor$.

Next it is argued \cite{bea72,dul92} that each fluid flows in
flow channels given by the solid matrix and the presence of the
other fluid. 
One postulates that to each saturation there corresponds
a unique configuration of flow channels.
This picture is then formalized into the constitutive
assumption that the relative permeabilities
\begin{eqnarray}
\kwr(\XX,\TT) & = & \kwr(\SS_n(\XX,\TT))\\
\kor(\XX,\TT) & = & \kor(\SS_n(\XX,\TT))
\label{Em7}
\end{eqnarray}
are functions of saturation alone.

Although these equations are widely accepted and almost
universally applied in numerical reservoir simulation \cite{tra88}
or aquifer modeling \cite{hel97}
they must be considered physically incomplete.
Even if the picture of flow channels is accepted it is
clear that the resulting permeability depends not only
on saturation (i.e. volume fraction) but also on the
surface area with other fluid or solid phases.
This experimental fact is well known from
observations of single phase flow in porous media
\cite{bea72,dul92}

The second basic problem with the macroscopic equations of motion
(\ref{Em1})-(\ref{Em7}) arises from the experimental
observation that the parameters $\Sor$ and $\Swi$
are neither constant nor are they known in advance
for a given experiment.
Instead they vary with space and time and depend on
the flow conditions in the experiment
\cite{DB54,tab69,TKS73,abr75,CKM88,WM85,LDS81,MCT85,OBP92}.
More precisely, the residual oil saturation depends
on the microscopic capillary number
$\Ca=\mu_\wa v/\sigma_{\oi\wa}$ where $v$ is a
typical flow velocity and $\sigma_{\oi\wa}$ is
the surface tension between the two fluids.
The dependence $\Sor(\Ca)$ is called
a capillary desaturation curve.
The measured values of $\Sor$ are known to depend
also on the duration of the flood.
Capillary desaturation experiments contradict clearly to the assumption that
the functions $\kwr(\SS_\wa),\kor(\SS_\wa)$ and $\PP_c(\SS_\wa)$
depend only on saturation.
Instead they depend also on velocity \cite{deg88,kal92} and pressure.
Hence they depend on the solution and cannot be considered to be 
constitutive relations characterizing the system.
Therefore the system of equations of motion
must be considered to be incomplete.



\section{Macroscopic conservation laws}

The basic idea of this paper is to include the energy
balance involving the macroscopic interfacial energy 
into the basic equations of motion.
As the interfacial tensions do not appear explicitly in equations
(\ref{Em1}) through (\ref{Em7}) it is clear that
this requires an enlarged description of the macroscopic state.
While previously the macroscopic state of the fluids was described
by saturations, pressures and velocities it must now include
macroscopic interfacial areas per unit volume.
The necessity to include interfacial areas in a macroscopic
description is also stressed by volume averaging approaches
\cite{mar82,whi86b,pav89}.
The main difference between volume averaging 
techniques and the approach of the present paper is that the
present approach is based purely on macroscopic mixture theory, 
and makes no reference to the underlying microscopic equations.

To discuss mass, energy and momentum balance laws in a unified 
manner consider first
a general conserved quantity $\EG$ defined per unit mass.
Let $\Flu$ denote the flux of this conserved quantity
across a surface of a material control region $\GGG(t)$ in $\RRR^3$.
The general multiphase conservation law in the theory of
mixtures reads in differential form \cite{all88,tru84}
\begin{equation}
\frac{D^\pha}{Dt}(\ph_\pha\rho_\pha\EG_\pha) + 
\ph_\pha\rho_\pha\EG_\pha\NAB\cdot \VV_\pha -
\NAB\cdot\Flu_\pha - \ph_\pha\rho_\pha\Ext_\pha = \Exc_\pha
\label{con1}
\end{equation}
Here the subscript $\alpha$ refers to the different phases
of the mixture, and
\begin{equation}
D^\pha/Dt=\partial/\partial t + \VV_\pha\cdot\NAB
\end{equation}
denotes the material derivative and 
$\VV_\pha$ is the velocity.
$\rho_\pha$ is the mass density, and $\ph_\pha$ 
is the volume fraction of phase $\alpha$.
Of course
\begin{equation}
\sum_\pha \ph_\pha = 1
\label{con2}
\end{equation}
and introducing the volume fraction of the void space,
i.e. the porosity,
\begin{equation}
\ph = \sum_{\pha\neq{\rm solid}} \ph_\pha
\end{equation}
allows to define saturations $\SS_\pha$ of the fluid phases through
\begin{equation}
\ph_\pha = \ph \SS_\pha
\label{con3}
\end{equation}
In (\ref{con1}) 
$\Ext_\pha$ denotes the external supply of $\EG$.
$\Exc_\pha$ represents the interphase transfer of the
conserved quantity $\EG$ into phase $\alpha$ from all
other phases. In other words
\begin{equation}
\Exc_\pha=\sum_{\phb}\Exc_{\pha\phb}
\end{equation}
where $\Exc_{\pha\phb}$ is the transfer of $\EG$ from phase $\beta$
into phase $\alpha$.
Of course global conservation of $\EG$ would require that summation
of (\ref{con1}) gives zero, $\sum_\pha \Exc_\pha = 0$, but
this will not be required here.
This general formalism can now be applied to mass, momentum
and energy conservation.

For mass conservation $\EG_\pha=1,\Flu_\pha=0,\Ext_\pha=0$ and 
$\Exc_\pha=\MT_\pha$ is the transfer of mass from all other
phases into phase $\alpha$.
Inserting into eq. (\ref{con1}) yields
\begin{equation}
\frac{\partial}{\partial t}(\ph_\pha\rho_\pha) + 
\NAB\cdot (\ph_\pha\rho_\pha\VV_\pha) 
= \MT_\pha
\label{con5}
\end{equation}
as usual.

For momentum conservation $\EG_\pha=\VV_\pha$ is the velocity,
$\Flu_\pha=\ST_\pha$ is the stress tensor,
$\Ext_\pha=\BF_\pha$ is the body force, and
$\Exc_\pha=\MMT_\pha$ is the momentum transfer from all other
phases into phase $\alpha$.
Inserting into (\ref{con1}) and using mass balance now yields
\begin{equation}
\ph_\pha\rho_\pha\frac{D^\pha\VV_\pha}{Dt} -
\NAB\cdot\ST_\pha - \ph_\pha\rho_\pha\BF_\pha = \MMT_\pha - \VV_\pha\MT_\pha
\label{con6}
\end{equation}
where the divergence $\NAB\cdot\ST_\pha$ of the stress tensor
is a vector whose $i$-th coordinate is 
$\sum_j\partial(\ST_\pha)_{ij}/\partial x_j$.

For energy conservation only the kinetic energy and the 
interfacial energies are considered, while thermal effects
will be ignored.
Hence in this case
\begin{equation}
\EG_\pha = \frac{1}{2}\VV_\pha\cdot\VV_\pha + 
\frac{1}{2}\sum_\phb\frac{\OS_{\pha\phb}\SIS_{\pha\phb}}
{\ph_\pha \rho_\pha+\ph_\phb \rho_\phb}
\label{con7}
\end{equation}
where $\OS_{\pha\phb}$ is the surface tension between phase $\alpha$ and
$\beta$ and $\SIS_{\pha\phb}$ is the specific internal surface
of the $\alpha\beta$-interface.
The denominator in the interfacial energy terms
(giving the $\alpha\beta$-density)
ensures the same normalization with respect
to mass and volume as for the kinetic energy term.
The factor $1/2$ in front of it attributes
the interfacial energy between phase $\alpha$ and $\beta$
equally to the two phases.
This assumption will be relaxed in a future publication
\cite{BH98}.
Proceeding with this assumption the energy flux becomes
\begin{equation}
\Flu_\pha = \ST_\pha\cdot\VV_\pha - \SEF_\pha
\label{con8}
\end{equation}
where $\ST_\pha\cdot\VV_\pha$ is the work done by stress
and $\SEF_\pha$ is the flux of surface energy.
The external energy supply is similarly split,
\begin{equation}
\Ext_\pha = \BF_\pha\cdot\VV_\pha + \SEE_\pha,
\label{con9}
\end{equation}
into the work $\BF_\pha\cdot\VV_\pha$ done by body forces
and the external supply $\SEE_\pha$ of interfacial energy.
Finally the energy exchange from all phases into phase 
$\alpha$ is $\Exc_\pha=\EEx_\pha$.

Inserting equations (\ref{con7}) through (\ref{con9}) into the general
balance law (\ref{con1}) gives the energy balance for immiscible
displacement as
\begin{eqnarray}
& &\frac{1}{2}\left(\NAB\cdot\VV_\pha+\frac{D^\pha}{Dt}\right) 
\left(\ph_\pha\rho_\pha \sum_\phb
\frac{\OS_{\pha\phb}\SIS_{\pha\phb}}
{\ph_\pha\rho_\pha+\ph_\phb\rho_\phb}\right)
+ \NAB\cdot\SEF_\pha - \ST_\pha:\NAB\VV_\pha -\ph_\pha\rho_\pha\SEE_\pha=
\nonumber\\[8pt] 
&=&\EEx_\pha- \MMT_\pha\cdot\VV_\pha
+ \MT_\pha\frac{\VV_\pha^2}{2}
\label{con10}
\end{eqnarray}
where
\begin{equation}
\ST_\pha:\NAB\VV_\pha=\sum_{i=1}^d\sum_{j=1}^d\left(\ST_\pha\right)_{ij}
\frac{\partial(\VV_\pha)_j}{\partial x_i}
\label{con11}
\end{equation}
Equations (\ref{con7}) and (\ref{con10}) embody the idea to
include interfacial areas and energies into the macroscopic
energy balance.
This is the first step beyond the traditional approach 
described e.g. in \cite{dul92,bea72,all88}.

The second step beyond the traditional theory consists in
splitting each fluid phase into a connected (percolating)
and disconnected (nonpercolating) subphase.
The idea here is that such a division is necessary to handle
the dynamics of residual phase saturations.
During an immiscible displacement process a finite fraction
of each fluid phase becomes trapped in the pores.
Once trapped the droplets are immobile and fixed through
large local capillary forces.
Only at very high flow rates can such trapped droplets be
mobilized.
Therefore the second main idea of the present theory is
to treat these droplets as an individual phase interacting
with the connected mother phase mainly via mass exchange.

Consider now the simplest case of two immiscible fluids
(called oil and water as before) flowing inside a rigid
porous medium.
Dividing each of the two fluid phases into two
subphases one has a total of five phases. 
Each fluid superphase is divided into a percolating 
(connected) and a trapped (disconnected) subphase. 
Here percolating means that each point in the subset 
occupied by the percolating subphase can be connected to
the exterior boundaries of the sample by a path within
the subphase.
The disconnected subphase is the complement within the 
chosen superphase.
The connected (percolating) subphase can also be defined as 
the region inside of which an external applied pressure
gradient can propagate.
Therefore the phase index $\alpha$ above can assume five values
summarized in Table \ref{table}.
\begin{table}
\caption{Overview over the subphases and their indices for
immiscible displacement of two fluids inside a rigid porous medium.}
\vspace*{1cm}
\begin{tabular}{c|c}
Phase index $\alpha$ & Phase description\\
\tableline\tableline
1 & connected (percolating) water phase\\
2  & disconnected (trapped) water phase\\
3  & connected (percolating) oil phase\\
4  & disconnected (trapped) oil phase\\
${\cal R}$ & rigid porous rock phase
\end{tabular}
\label{table}
\end{table}

\section{Constitutive relations and constraints}

This section complements the basic balance laws with geometrical
and physical constraints and with constitutive relations.

The fluids are assumed to be incompressible
\begin{eqnarray}
\rho_\Wc(\XX,t) & = & \rho_\Wd(\XX,t) = \rho_\wa = {\rm const}\\
\rho_\Oc(\XX,t) & = & \rho_\Od(\XX,t) = \rho_\oi = {\rm const}
\label{ass1}
\end{eqnarray}
and their stress tensors are approximated as
\begin{equation}
\ST_\pha = -\ph_\pha\PP_\pha\;\one
\label{ass2}
\end{equation}
for all $\alpha\in\{1,2,3,4\}$ where $\one$ is the unit tensor.
In (\ref{ass2}) it is assumed that all shear stresses
in the fluid are negligible, and that $\PP_\pha$ are the
true pressures in the fluids.
It will be assumed that the only body force is gravity,
$\BF_\pha=g\:\NAB\ZZ$ where $g$ is the acceleration of
gravity and $\ZZ(\XX)$ is the depth function.
In addition inertial effects will be neglected,
$D^\pha\VV_\pha/Dt = 0$, for all phases.
All of these assumptions are also made in the 
derivation of the traditional equations described in the 
previous section \cite{dul92,bea72,all88}.

It will be assumed that there are no external sources or sinks
of surface energy, $\Ext_\pha=0$.
Next it is assumed that $\SEF_\pha=0$, i.e. there is no diffusive
flux of surface energy during the time scale of observation.
A crucial assumption is that the exchange of interfacial
energy between the phases is dominated by separation and
coalescence of disconnected subphase droplets from/with 
their connected mother phase, i.e. through mass transfer, 
and that all other mechanisms are negligible.
This amounts to setting $\EEx=0$.

The porous matrix as well as the disconnected (trapped) subphases 
are immobile, and hence
\begin{equation}
\VV_\ro = \VV_\Wd = \VV_\Od = 0 .
\label{ass3}
\end{equation}
The interfacial area between a disconnected subphase and its
superphase vanishes, and the same holds for the contact between
the two disconnected subphases. Hence
\begin{equation}
\SIS_{\Wc\Wd} = \SIS_{\Oc\Od} = \SIS_{\Wd\Od} = 0
\label{ass4}
\end{equation}
Of course $\SIS_{\pha\pha}=0$ and symmetry
$\SIS_{\pha\phb}=\SIS_{\phb\pha}$ holds.

The disconnected fluid phases have macroscopic pressures
$\PP_\Wd(\XX,t)$ and $\PP_\Od(\XX,t)$
obtained by averaging the pressure in the disconnected 
drops and droplets.
The resulting functions $\PP_\Wd(\XX,t),\PP_\Od(\XX,t)$
however are generally not continuous and hence not
differentiable.

The wetting properties are assumed to be constant, and they are
specified by the requirements that the individual interface areas 
with the rock are constant, i.e.
\begin{eqnarray}
\SIS_{\Wc\ro}(\XX,t) & = & \SIS_{\Wc\ro} = {\rm const}\label{ass5a}\\
\SIS_{\Wd\ro}(\XX,t) & = & \SIS_{\Wd\ro} = {\rm const}\\
\SIS_{\Oc\ro}(\XX,t) & = & \SIS_{\Oc\ro} = {\rm const}\\
\SIS_{\Od\ro}(\XX,t) & = & \SIS_{\Od\ro} = {\rm const} .\label{ass5}
\end{eqnarray}
Of course $\SIS=\SIS_{\Wc\ro}+\SIS_{\Wd\ro}+\SIS_{\Oc\ro}+\SIS_{\Od\ro}$
is the total specific internal surface of the porous medium.
The ratio
\begin{equation}
\WI_\wa =\frac{\SIS_{\Wc\ro}+\SIS_{\Wd\ro}}{\SIS}
\label{ass6}
\end{equation}
varies from $1$ to $0$ as the medium varies from water wet to oil wet.

Finally it remains to specify the momentum exchanges 
$\MMT_\pha = \sum_\phb \MMT_{\pha\phb}$ and
the mass exchanges $\MT_\pha = \sum_\phb \MT_{\pha\phb}$.
In both cases 
$\MMT_{\pha\pha}=0, \MT_{\pha\pha}=0$ will be used.
The momentum exchange is conventionally modelled \cite{all88}
through Stokes drag in the form
\begin{equation}
\MMT_{\pha\phb} = \wid_{\pha\phb}(\VV_\phb-\VV_\pha) =
\frac{\ph_\pha}{\Mob_{\pha\phb}}(\VV_\phb-\VV_\pha)
\label{ass7}
\end{equation}
where the mobility $\Mob_{\pha\phb}=\KK_{\pha\phb}/\mu_{\pha\phb}$
is expressed as usual in terms of the permeability $\KK_{\pha\phb}$
and the viscosity $\mu_{\pha\phb}$.
As a new assumption the mobilities $\Mob_{\pha\phb}$ are modeled 
through
\begin{equation}
\Mob_{\pha\phb} = \frac{\KK_{\pha\phb}}{\mu_{\pha\phb}} = 
\frac{\ph_\pha^3}{\mu_\pha\SIS^2_{\pha\phb}}
\label{ass8}
\end{equation}
analogous to the Carman-Kozeny relation for absolute
permeabilities.
This assumption reflects the traditional view that
each fluid flows in its own channel.
If there is no common interface, then it is not possible
to exchange momentum, hence
\begin{equation}
\MMT_{\Wc\Wd} = \MMT_{\Oc\Od} = \MMT_{\Wd\Od} = 0
\label{ass9}
\end{equation}
completes the specification of the momentum exchange.

The rigid porous matrix has a volume fraction $1-\ph=$const which
is assumed to be spatially and temporally constant.
The rigid matrix is dynamically inert (see eq.(\ref{ass3})), and hence 
only the balance laws for the fluid phases will appear below.
\footnote{\label{foot}Because the rock matrix is assumed to be rigid
and impermeable one has
$\wid_{\ro\Wc} = \wid_{\ro\Wd} = \wid_{\ro\Oc} =
\wid_{\ro\Od}=\infty$ 
which implies dissipation.
Because of this and (\ref{ass3}) the balance laws for
the matrix phase would either be trivial or ill defined.}

Chemical reactions between oil, water and rock are excluded, and hence
\begin{equation}
\MT_{\Wc\Oc} = \MT_{\Wc\Od} = \MT_{\Wd\Oc} = \MT_{\Wd\Od} =
\MT_{\Wc\ro} = \MT_{\Wd\ro} = \MT_{\Oc\ro} = \MT_{\Od\ro} = 0 .
\end{equation}
Mass transfer occurs within each fluid phase between its 
connected and the disconnected subphases.
These nonvanishing mass transfers are
\begin{eqnarray}
\MT_\wa = \MT_{\Wc\Wd} & = & -\MT{\Wd\Wc}\\
\MT_\oi = \MT_{\Oc\Od} & = & -\MT{\Od\Oc}
\end{eqnarray}
In general the mass transfer may be a function of the velocities,
saturations and interface areas of the other phases.
A precise form for the dependence 
$\MT_\wa = \MT_\wa(\VV_\Wc,\SIS_{\Wc\Oc},\SS_\Wc \SS_\Oc)$
is not needed here. 
Any concrete assumption for such a dependence would have to 
be checked by experiment.

\section{Results}
Combining the balance laws with the constitutive equations gives
the macroscopic equations of motion of the two pore fluids.
\footnote{Balance equations for the rock matrix do 
not arise (see footnote \ref{foot}).}
Mass conservation yields four equations
\begin{eqnarray}
&\rho_\wa&\ph\frac{\partial}{\partial t}\SS_\Wc + 
\rho_\wa\ph\NAB\cdot (\SS_\Wc\VV_\Wc) = \MT_\wa
\label{res1}\\
&\rho_\wa&\ph\frac{\partial}{\partial t}\SS_\Wd = -\MT_\wa
\label{res2}\\
&\rho_\oi&\ph\frac{\partial}{\partial t}\SS_\Oc + 
\rho_\oi\ph\NAB\cdot (\SS_\Oc\VV_\Oc) = \MT_\oi
\label{res3}\\
&\rho_\oi&\ph\frac{\partial}{\partial t}\SS_\Od = -\MT_\oi
\label{res4}
\end{eqnarray}
while momentum conservation gives only two equations,
\begin{eqnarray}
\NAB(\ph_\Wc\PP_\Wc)-\ph_\Wc\rho_\wa g\:\NAB\ZZ & = &
\wid_{\Wc\Oc}\VV_\Oc - 
(\wid_{\Wc\Oc}+\wid_{\Wc\Od}+\wid_{\Wc\ro}+\MT_\wa)\VV_\Wc
\label{res5}\\
\NAB(\ph_\Oc\PP_\Oc)-\ph_\Oc\rho_\oi g\:\NAB\ZZ & = &
\wid_{\Oc\Wc}\VV_\Wc - 
(\wid_{\Oc\Wc}+\wid_{\Oc\Wd}+\wid_{\Oc\ro}+\MT_\oi)\VV_\Oc
\label{res6}
\end{eqnarray}
because $\PP_\Wd$ and $\PP_\Od$ are not differentiable.
Energy conservation gives again four equations
\begin{eqnarray}
&&\frac{1}{2}
\left(\NAB\cdot\VV_\Wc+\frac{D^\Wc}{Dt}\right)
\left(\frac{\ph_\Wc\rho_\wa\OS_{\wa\oi}\SIS_{\Wc\Oc}}
{\rho_\wa\ph_\Wc+\rho_\oi\ph_\Oc} +
\frac{\ph_\Wc\rho_\wa\OS_{\wa\oi}\SIS_{\Wc\Od}}
{\rho_\wa\ph_\Wc+\rho_\oi\ph_\Od} +
\frac{\ph_\Wc\rho_\wa\OS_{\wa\ro}\SIS_{\Wc\ro}}
{\rho_\wa\ph_\Wc+\rho_\ro(1-\ph)} \right)\nonumber\\[6pt]
&&\hspace*{1cm}= -\ph_\Wc\PP_\Wc\NAB\cdot\VV_\Wc 
-\wid_{\Wc\Oc}\VV_\Wc\cdot\VV_\Oc
+(\wid_{\Wc\Oc}+\wid_{\Wc\Od}+\wid_{\Wc\ro})\VV^2_\Wc
+\MT_\wa\frac{\VV^2_\Wc}{2}
\label{res7}\\[10pt]
&&\frac{\partial}{\partial t}
\left(\frac{\ph_\Wd\rho_\wa\OS_{\wa\oi}\SIS_{\Wd\Oc}}
{\rho_\wa\ph_\Wd+\rho_\oi\ph_\Oc} +
\frac{\ph_\Wd\rho_\wa\OS_{\wa\ro}\SIS_{\Wd\ro}}
{\rho_\wa\ph_\Wd+\rho_\ro(1-\ph)}\right)= 0\label{res8}\\[10pt]
&&\frac{1}{2}
\left(\NAB\cdot\VV_\Oc+\frac{D^\Oc}{Dt}\right)
\left(\frac{\ph_\Oc\rho_\oi\OS_{\wa\oi}\SIS_{\Wc\Oc}}
{\rho_\wa\ph_\Wc+\rho_\oi\ph_\Oc} +
\frac{\ph_\Oc\rho_\oi\OS_{\wa\oi}\SIS_{\Wd\Oc}}
{\rho_\wa\ph_\Wd+\rho_\oi\ph_\Oc} +
\frac{\ph_\Oc\rho_\oi\OS_{\oi\ro}\SIS_{\Oc\ro}}
{\rho_\oi\ph_\Oc+\rho_\ro(1-\ph)} \right)\nonumber\\[6pt]
&&\hspace*{1cm}= -\ph_\Oc\PP_\Oc\NAB\cdot\VV_\Oc
-\wid_{\Oc\Wc}\VV_\Wc\cdot\VV_\Oc
+(\wid_{\Oc\Wc}+\wid_{\Oc\Wd}+\wid_{\Oc\ro})\VV^2_\Oc
+\MT_\oi\frac{\VV^2_\Oc}{2}
\label{res9}\\[10pt]
&&\frac{\partial}{\partial t}
\left(\frac{\ph_\Od\rho_\oi\OS_{\wa\oi}\SIS_{\Wc\Od}}
{\rho_\wa\ph_\Wc+\rho_\oi\ph_\Od} +
\frac{\ph_\Od\rho_\oi\OS_{\oi\ro}\SIS_{\Od\ro}}
{\rho_\oi\ph_\Od+\rho_\ro(1-\ph)}\right) = 0
\label{res10}
\end{eqnarray}
The unknown fields in these equations are the four
saturations $\SS_\pha$ ($\alpha=1,2,3,4$) of the fluids related 
to $\ph_\pha$ by equation (\ref{con3}),
the two pressures $\PP_\Wc,\PP_\Oc$, the six components of
the two velocities $\VV_\Wc,\VV_\Oc$ of the connected fluids, 
and the three surface areas 
$\SIS_{\Wc\Oc},\SIS_{\Wc\Od},\SIS_{\Wd\Oc}$.
Note that all $\SIS_{\pha,\ro}, \alpha=1,2,3,4$, are
constant by virtue of eqs. (\ref{ass5a}) through (\ref{ass5}).
This is a total of 15 unknowns for which 
the condition $\SS_\Wc+\SS_\Wd+\SS_\Oc+\SS_\Od=1$
(see (\ref{con2})) together with equations (\ref{res1})--(\ref{res10})
form a set of 15 equations.

A detailed analysis of these equations will be presented elsewhere
\cite{BH98}.
Here only the simplest comparison with the traditional
equations (\ref{Em1})--(\ref{Em7}) will be discussed.
In particular the question arises whether the new equations
allow to determine relative permeabilities and capillary
pressure functions.

The capillary pressure, defined in eq. (\ref{Em4}),
is an equilibrium property. 
In the traditional equations it is related to the 
saturation profile in complete gravitational 
equilibrium \cite{bea72}, i.e. when
$\VV_\pha=0$ and when all
terms involving time derivatives are set to zero.
Proceeding in this way one finds that the resulting equations
can be solved to give
\begin{equation}
\PP_c(\XX,\infty) = \PP_\Oc - \PP_\Wc =
\frac{c_\Oc}{\ph\SS_\Oc(\XX,\infty)} -
\frac{c_\Wc}{\ph\SS_\Wc(\XX,\infty)}
+(\rho_\oi-\rho_\wa)\XX\cdot\GG
\label{res11}
\end{equation}
where $\GG$ is the acceleration vector of gravity 
and $c_\Oc,c_\Wc$ are integration constants.
Using $\SS_\wa=\SS_\Wc+\SS_\Wd$ and 
$1-\SS_\wa=\SS_\Oc+\SS_\Od$
shows that in the directions perpendicular to gravity
\begin{equation}
\PP_c(\XX,\infty) = \PP_\Oc - \PP_\Wc =
\frac{c_\Oc}{\ph(1-\SS_\Od-\SS_\wa)} -
\frac{c_\Wc}{\ph(\SS_\wa-\SS_\Wd)}
\label{res12}
\end{equation}
If it is assumed that $\SS_\Wd$ and $\SS_\Od$ are 
spatially constant throughout the sample then it is
tempting to identify them with $\Swi$ and $\Sor$
because this would imply the assumption (\ref{Em4}) 
of the traditional formulation.
Moreover the capillary pressure would indeed have
the familiar S-shaped form as a function of
saturation.
It should be noticed however that eq.(\ref{res12})
holds only without flow while assumption (\ref{Em4})
is assumed to hold always.
Therefore in the new formulation the capillary pressure
depends strongly on the flow regime.
This prediction agrees with what is known from
theory and experiment \cite{deg88,kal92}.
Notice also that in (\ref{res12}) the quantities
$\SS_\Wd$ and $\SS_\Od$ are state variables
and hence the capillary pressure includes 
the spatiotemporal variability of the residual
saturations.

The generalization of Darcy's law follows 
from the equations of
momentum balance (\ref{res5}) and (\ref{res6})
by solving these equations for $\VV_\Wc$
and $\VV_\Oc$.
This yields a generalized form of Darcy's law
for two phase flow in the form
\begin{eqnarray}
\VV_\Wc &=& \kr_{\wa\wa}\frac{k}{\mu_\wa}
(\NAB(\ph_\Wc\PP_\Wc)-\ph_\Wc\rho_\wa\GG)
+\kr_{\wa\oi}\frac{k}{\mu_\oi}
(\NAB(\ph_\Oc\PP_\Oc)-\ph_\Oc\rho_\oi\GG)
\label{res13}\\
\VV_\Oc &=& \kr_{\oi\wa}\frac{k}{\mu_\wa}
(\NAB(\ph_\Wc\PP_\Wc)-\ph_\Wc\rho_\wa\GG)
+\kr_{\oi\oi}\frac{k}{\mu_\oi}
(\NAB(\ph_\Oc\PP_\Oc)-\ph_\Oc\rho_\oi\GG)
\label{res14}
\end{eqnarray}
where the relative permabilities $\kr_{\pha\phb}$
are defined by these equations.
The $\kr_{\pha\phb}$ are related to the resitivities $\wid_{\pha\phb}$
which in turn are given by equations (\ref{ass7}) and (\ref{ass8}).

This result differs from the traditional generalization
of Darcy's law given in eqs. (\ref{Em2}) in three
respects. 
Firstly (\ref{res13}) and (\ref{res14}) includes
off-diagonal coupling terms $\kr_{\wa\oi}$ and $\kr_{\oi\wa}$
which are missing in (\ref{Em2}).
Secondly (\ref{res13}) and (\ref{res14}) include
a new driving force $\ph\PP_\pha\NAB\SS_\pha$
resulting from the term
$\NAB(\ph_\pha\PP_\pha)$.
The new driving force is proportional to saturation gradients.
Thirdly, and perhaps most importantly, the relative
permeabilities defined through (\ref{res13}) and (\ref{res14})
depend not only on saturation as assumed in (\ref{Em7})
but 
\begin{eqnarray}
\kr_{\pha\phb} & = &
\kr_{\pha\phb}(\wid_{\Wc\Oc},\wid_{\Oc\Wc},\wid_{\Wc\Od},\wid_{\Wd\Oc},
\wid_{\Wc\ro},\wid_{\Oc\ro},\MT_\wa,\MT_\oi)\\[6pt]
& = & \kr_{\pha\phb}
(\SS_\Wc,\SS_\Wd,\SS_\Oc,\SS_\Od,\SIS_{\Wc\Oc},\SIS_{\Wc\Od},\SIS_{\Oc\Wd},
\partial\SS_\Wd/\partial\TT,\partial\SS_\Od/\partial\TT,
\SIS_{\Wc\ro},\SIS_{\Oc\ro})
\label{res15}
\end{eqnarray}
by virtue of (\ref{res5}),(\ref{res6}) and
(\ref{ass7}),(\ref{ass8}),(\ref{res2}) and (\ref{res4}).
This expresses a dependence on saturation, changes of saturation,
disconnected (residual) saturations, interfacial area and 
wetting properties ($\SIS_{\Wc\ro},\SIS_{\Oc\ro}$).
Thus the present theory gives a much more detailed
picture of relative permeabilities than the traditional
assumption (\ref{Em7}).

In conclusion the present paper has discussed the problems
associated with the macroscopic equations of motion for two
phase immiscible displacement in porous media.
After discussion of the traditional formulation a new
theory has been developed based on mixture theory.
The new equations differ from the traditional ones
in the following main aspects:
\begin{enumerate}
\item
The set of macroscopic state variables is
enlarged to include specific surface area of
the fluid phases.
\item 
The set of macroscopic balance laws is
enlarged to include interfacial energy balance.
\item 
The theory allows a macroscopic quantification
of the wetting properties through the surface areas.
\item 
The dynamics of the disconnected fluid phases
is included.
This allows to describe the spatiotemporal 
behaviour of residual saturations.
\item 
The theory predicts a dynamic capillary
pressure which, at least in complete gravitational
equilibrium, shows the correct qualitative
behaviour as a function of saturation.
\item 
The relative permeabilities are found to depend
on macroscopic state variables other than saturation.
In particular they depend on wetting properties
and the internal surfaces areas.
\end{enumerate}
All of these results are consistent with experiment.
More analysis of the equations is necessary to
compare their solutions with experiment.

\vspace*{2cm}\noindent
ACKNOWLEDGEMENT\\
I am grateful to
Dr. H. Besserer for discussions and for
critically reading the manuscript, and to the Deutsche 
Forschungsgemeinschaft for financial support.

\end{document}